\documentclass[12pt]{article}
\newcommand{\be}{\begin{equation}}
\newcommand{\bea}{\begin{eqnarray}}
\newcommand{\eea}{\end{eqnarray}}
\newcommand{\ba}{\begin{array}}
\newcommand{\ea}{\end{array}}
\newcommand{\ee}{\end{equation}}

\expandafter\ifx\csname mathbbm\endcsname\relax

\else

\fi
\begin{document}
\begin{titlepage}
\hfill \vbox{
    \halign{#\hfil         \cr
           IPM/P-2003/045 \cr
           ITFA-2003-38\cr
           hep-th/0308120  \cr
           } 
      }  
\vspace*{16mm}
\begin{center}
{\Large {\bf ${\cal N}=1\; G_2$ SYM theory  and Compactification
to Three Dimensions}\\ }

\vspace*{15mm} \vspace*{1mm} {Mohsen Alishahiha$^a$
\footnote{Alishah@theory.ipm.ac.ir}, Jan de Boer$^b$
\footnote{jdeboer@science.uva.nl}, Amir E. Mosaffa$^{a,c}$}
\footnote{Mosaffa@theory.ipm.ac.ir} and Jeroen Wijnhout$^b$
\footnote{wijnhout@science.uva.nl}
 \\
\vspace*{8mm}

{\it$^a$ Institute for Studies in Theoretical Physics
and Mathematics (IPM)\\
P.O. Box 19395-5531, Tehran, Iran \\ \vspace{3mm}
$^b$ Instituut voor Theoretische Fysica,\\
Valckenierstraat 65, 1018XE Amsterdam, The Netherlands\\
\vspace{3mm}
$^c$ Department of Physics, Sharif University of Technology\\
P.O. Box 11365-9161, Tehran, Iran}\\

\vspace*{8mm}
\end{center}

\begin{abstract}
We study four dimensional ${\cal N}=2$ $G_2$ supersymmetric gauge
theory on $R^3\times S^1$ deformed by a tree level superpotential.
We will show that the exact superpotential can be obtained by
making use of the Lax matrix of the corresponding integrable model
which is the periodic Toda lattice based on the dual of the affine
$G_2$ Lie algebra. At extrema of the superpotential the
Seiberg-Witten curve typically factorizes, and we study the
algebraic equations underlying this factorization. For $U(N)$
theories the factorization was closely related to the geometrical
engineering of such gauge theories and to matrix model
descriptions, but here we will find that the geometrical
interpretation is more mysterious. Along the way we give a method
to compute the gauge theory resolvent and a suitable set of
one-forms on the Seiberg-Witten curve. We will also find evidence
that the low-energy dynamics of $G_2$ gauge theories can
effectively be described in terms of an auxiliary hyperelliptic
curve.

\end{abstract}

\end{titlepage}

\section{Introduction}
Supersymmetric gauge theories have been in the center of attention
for a long time. One of the reasons for this is that a large class
of these theories, {\it i.e.} ${\cal N}=1$ gauge theories, are
likely to be of relevance for real world physics. The other reason
is that non-supersymmetric gauge theories such as ordinary QCD can
be considered as perturbations away from a supersymmetric point.

These theories have a rich structure and one can obtain exact
results about their non-perturbative dynamics and hence about
their vacuum structure (for a review, see e.g
\cite{Intriligator:1995au}). A major step towards the
understanding of this structure was a general organizing principle
put forward by Dijkgraaf and Vafa \cite{Dijkgraaf:2002dh}.
Motivated by earlier works
\cite{Bershadsky:1993cx}-\cite{Dijkgraaf:2002vw} these authors
have conjectured that the exact superpotential and gauge couplings
for a wide class of ${\cal N}=1$ gauge theories can be obtained by
doing perturbative calculations in a dual matrix model only taking
the planar diagrams into account. This conjecture has been
verified using perturbative superspace techniques
\cite{Dijkgraaf:2002xd}, and also using anomalies
\cite{Cachazo:2002ry}.

Despite the successes of this conjecture some of its features
remain somewhat puzzling such as the distinguished role of the
gluino condensate superfields, the appearance of the
Veneziano-Yankielowicz superpotential and the capability of the
matrix model approach to include all possible gauge theories.

In trying to answer some of these questions, ${\cal N}=2\; U(N)$
theories deformed by a ${\rm Tr} W(\phi)$ superpotential were
considered on the space $R^3\times S^1$ \footnote{The
compactification of the ${\cal N}=2$ SYM theory to three
dimensions was considered in \cite{Seiberg:1996nz}. For further
discussions see for example
\cite{Katz:1996th}-\cite{Aharony:1997bx}.} in \cite{Boels:2003fh}
where, based on earlier works \cite{Dorey:1999sj}, it was
conjectured that if the classical superpotential is ${\rm Tr}
W(\phi)$ then the quantum superpotential will be just ${\rm Tr }
W(M)$ where $M$ is the Lax matrix of the integrable system that
underlies the four dimensional theory.\footnote{The relation
between ${\cal N}=2$ SYM theories and integrable system was
discussed in several papers including
\cite{Gorsky:1995zq}-\cite{Mironov:2000se}. For recent discussion
in this direction and its relation with Dijkgraaf-Vafa conjecture
see also \cite{Itoyama:2002rk}.} In a consequent publication
\cite{Boels:2003at} the agreement of the vacuum structure obtained
by the Lax matrix approach with the results obtained in four
dimensions using the conventional field theoretic approach was
proved for the gauge group $U(N)$. In a separate work
\cite{Hollowood:2003ds} the same result was proved using
alternative methods. The above conjecture was tested for gauge
groups $SO/SP$ in \cite{Alishahiha:2003pu} and again a complete
agreement with the known results was shown. For some related
comments see also \cite{Aganagic:2003xq}.

If we replace classical groups by exceptional groups, several new
questions arise. In \cite{Aganagic:2003xq} it was shown that the
perturbative computation of the glueball superpotential described
in \cite{Dijkgraaf:2002xd} reduces to effectively zero-dimensional
integrals even for exceptional groups. Therefore, one would expect
that there still exists an appropriate notion of a matrix
integral, however the meaning of the ``planar diagrams'' in such a
matrix theory is not clear, nor is it known what replaces the
Calabi-Yau geometry that was used to solve the matrix theory for
the classical groups. Another issue is related to the ambiguity of
the glueball superpotential also discussed in
\cite{Aganagic:2003xq}. This ambiguity arises because the gauge
theory with an arbitrary superpotential is not renormalizable, and
to make it into a well-defined theory one needs to specify a
suitable UV completion. This can in general be done in different
ways, leading to different answers for terms of sufficiently high
order in the glueball superfield. String theory prefers in some
sense one particular UV completion, and a natural field-theoretic
UV completion using embeddings in supergroups was described in
\cite{Aganagic:2003xq}. This latter technique fails for
exceptional groups, for which no natural field-theoretic UV
completion is known.

To study these questions we consider in this paper the example of
$G_2$, the exceptional group of lowest rank. We will find partial
answers to the above questions. In particular, we will find some
algebraic equations that in principle determine the geometry
underlying the exceptional matrix models, though we were not able
to put them in a nice form. We will also see that the Lax matrices
provide a natural UV completion of gauge theories for all gauge
groups, including exceptional ones, at least as far as holomorphic
quantities are concerned.

The paper is organized as follows. In section 2 the classical
description of ${\cal N}=2$ SYM theory with gauge group $G_2$ is
considered. Then the deformation by a tree level potential is
studied. In section 3 the quantum description of ${\cal N}=1$
$G_2$ four dimensional SYM is studied and the effective
superpotential is computed.\footnote{${\cal N}=1$ $G_2$ four dimensional 
SYM coupled to the different matters is also studied
\cite{{Pesando:1995bq},{Brandhuber:2003va}}.}  The Lax matrix of the related
integrable system is introduced and the different configurations
where the Seiberg-Witten curve is factorized  corresponding to the
unbroken and broken gauge group cases as well as the
superconformal field theory case are studied. In section 4 the
theory on $R^3\times S^1$ is considered and again the unbroken
gauge symmetry case and the broken one are studied. In the former
case the effective action in terms of the glueball field is
obtained.The vacuum solutions are also interpreted from the view
point of the corresponding integrable model. In section 5 we
describe how the Lax matrix provides a UV completion and derive
the corresponding gauge theory resolvent. We also consider the
factorization problem and give a proof that enables us to state
the extremization problem in purely algebraic terms. These
equations should describe the geometry underlying the exceptional
matrix model. In section 6 an argument is presented supporting the
existence of a hyperelliptic curve for $G_2$. The last section is
devoted to conclusion and remarks.

\section{ Classical description}
In this section we review some classical aspects of the ${\cal N}=2$ SYM theory
with gauge group $G_2$. The theory has a Coulomb branch where the gauge
group is
broken to
$U(1)^2$. The classical moduli space of the Coulomb branch is described by the
characteristic polynomial
\be
{\cal P}_{\rm class}(x)={1\over x}\det(x{\bf 1}-\phi)=x^6-2ux^4+u^2x^2-v\;,
\label{ccurve}
\ee
where $\phi$ is the adjoint scalar component of the ${\cal N}=1$
chiral multiplet contained in the ${\cal N}=2$ vector multiplet.
We assume in (\ref{ccurve}) that it takes values in the
seven-dimensional fundamental representation of $G_2$, and using a
gauge transformation and the equations of motion it can be assumed
to take values in the Cartan subalgebra. The moduli parameters of
the polynomial $u$ and $v$ are defined in terms of the gauge
invariant parameters of the gauge group as follows
\be
u={1\over 2}u_2\;,\;\;\;\;\;v=u_6-{1\over 12}u_2^3\;,\;\;\;\;\;\;\;{\rm
where}\;\;
u_k={1\over k}\;{\rm Tr}(\phi^k)\;.
\ee
The classical discriminant of the polynomial (\ref{ccurve}) up to a redundant
numerical factor is
\be
\Delta_{\rm class}=-4u^3v+27v^2\;,
\ee
whose zeroes give the points on the classical moduli space where
the two of the zeroes of the polynomial coincide. Note that the
classical discriminant is invariant under the following duality
transformation
\be
v\rightarrow -v+{4\over 27}u^3\;,
\label{DU}
\ee
which reflects the fact that the root lattice of $G_2$ is self-dual.

Let us now consider a deformation of the theory given by adding a tree level
superpotential given by
\be
W_{\rm tree}=g_2u+g_6v\;.
\ee
This deformation lifts most of the classical moduli space. To have
a supersymmetric vacuum one needs to impose the D- and F-term
equations. Taking $\phi$ to be diagonal implies that the D-term
equation is satisfied, and for the F-term equations one should set
$W'=0$. More precisely, taking
\be
\phi_{\rm class}={\rm diag}(\phi_1+\phi_2,2\phi_1-\phi_2,\phi_1-2\phi_2,
2\phi_2-\phi_1, \phi_2-2\phi_1,
-\phi_1-\phi_2,0)
\ee
the F-term condition reads
\bea
(\phi_2-2\phi_1)\left(g_2+2g_6(2\phi_2^4+2\phi_1^4+5\phi_2^3\phi_1
-3\phi_2^2\phi_1^2-4\phi_2\phi_1^3)\right)&=&0\;,\cr &&\cr
(\phi_1-2\phi_2)\left(g_2+2g_6(2\phi_1^4+2\phi_2^4+5\phi_1^3\phi_2
-3\phi_1^2\phi_2^2-4\phi_1\phi_2^3)\right)&=&0\;,
\eea
which has two inequivalent solutions given by
\begin{equation}
\phi_1=\phi_2=0
\end{equation}
and
\begin{equation}
 \phi_1=\phi_2=(-g_2/4g_6)^{1/4}.
 \label{eq:clasSU2U1}
\end{equation}
The first one corresponds to the case
where the gauge group remains unbroken while the second one corresponds
to the situation where the gauge group is broken to $SU(2)\times U(1)$ and
in this case the classical superpotential is
$W_{\rm tree}=g_2(-g_2/4g_6)^{1/2}$.

Using the explicit form of the gauge invariant parameters for these solutions
one can see that the discriminant is also zero. Therefore the solutions
correspond to the situation where the classical curve becomes degenerate.
Explicitly one finds
\bea
& \phi_1=\phi_2=0 &\longrightarrow\;\;\;\;  {\cal P}_{\rm class}=x^6\cr
& \phi_1=\phi_2= e& \longrightarrow\;\;\;\;
{\cal P}_{\rm class}=(x^2-e^2)^2(x^2-4e^2)\;,
\eea
where $e=(-g_2/ 4g_6)^{1/4}$. Note that by making use of the
duality transformation one could get a degenerate curve of the
form $x^2(x^2-3e^2)^2$.

\section{Quantum description}

In this section we study the quantum aspects of ${\cal N}=1\; G_2$
four dimensional SYM theory. In particular we shall compute the effective
superpotential. In fact since our ${\cal N}=1$ theory can be
thought of as an ${\cal N}=2$ theory deformed by a superpotential,
one can use the exact result of ${\cal N}=2\; G_2$ SYM theory to compute
the quantum superpotential.
Actually, one might suspect that the exact superpotential can be
obtained from some kind of the Seiberg-Witten curve factorization,
though, in the case of $G_2$ the curve is not hyperelliptic.

The Seiberg-Witten curve of $G_2$ gauge theory is given by the
spectral curve of the periodic Toda chain based on the dual affine
Lie algebra $G_2^{(1)}$ which is given in terms of twisted affine
Lie algebra $D_4^{(3)}$ (see for example \cite[page
511]{vilenkin}). It is well known that the underlying integrable
system, the Toda chain, admits a Lax pair for arbitrary gauge
groups. Thus there exist two matrices $M$ and $A$ such that
evolution of the theory can be described by the Lax equation
\be
{\partial M\over \partial t}=[M,A]\;.
\ee
In our model the corresponding Lax matrix is given by
\bea
&&M=\cr &&\cr
&&\pmatrix{\phi_1+\phi_2&y_2& 0&0&0&y_1&-z&0\cr 1&2\phi_1-\phi_2
&0&ay_1 &by_1 &0 &0 &-z \cr
0&0 &\phi_1-2\phi_2 &-a &b &0 &0 &y_1 \cr 0&a &-ay_2 &0 &0 &-a &ay_1 &0 \cr
0&b &by_2 & 0&0 &b &by_1 &0 \cr 1&0 &0 &-ay_2 &by_2 &2\phi_2-\phi_1 &0 &0 \cr
-{y_0\over z}&0 &0 &a &b &0 &\phi_2-2\phi_1 &y_2 \cr 0&-{y_0\over z} &1
&0 &0 &0
 &1 &-\phi_1-\phi_2 },\nonumber
\eea
where $a=\sqrt{1/2},\;b=\sqrt{3/2}$ and there is a constraint on $y_i$
given by $y_0y_2y_1^2=\Lambda^{8}/36$. The Seiberg-Witten curve is then
obtained
from the spectral curve $\det(x{\bf 1}-M)=0$, which is
\be
{\cal P}_{\rm quan}(z,x):=3(z-{\Lambda^8\over
36z})^2-x^2(z+{\Lambda^8\over 36z})
(6x^2-2u)-x^2P(x,u,v)=0\;,
\label{G2q}
\ee
where
\be
P(x,u,v)=x^6-2ux^4+u^2x^2-v\;.
\ee
Here $u$ and $v$ are the moduli of the quantum curve which are functions of
$\phi_i$ and $y_i$. Therefore the quantum moduli space of the Coulomb branch
of $G_2$ theory is parameterized by $u$ and $v$. These parameters can be
given in terms of traces of the Lax matrix as follows
\be
u={1\over 2}U_2\;,\;\;\;\;\;v=U_6-{1\over 12}U_2^3+5U_2(z+{y_0y_1^2y_2\over z}),
\;\;\;\;\;{\rm where}\;\;U_k={1\over k}\;{\rm Tr}(M^k)\;.
\ee
The last term in the expression of $v$ is necessary because ${\rm
Tr}(M^6)$ appears to
depend explicitly on the spectral parameter $z$ and in order to remove
the $z$ dependence of $v$ one needs to have this extra term.

It is important to note that since the Seiberg-Witten curve is based on
the dual algebra, to compare our results with the field theory results one
should perform a duality transformation as (\ref{DU}). Therefore we will
consider the theory with the following tree level superpotential
\be
W_{\rm tree}=g_2u-g_6v+{4\over 27}g_6u^3\;.
\ee

To find the supersymmetric vacua and the corresponding quantum
superpotential we will need to consider the factorization of the
Seiberg-Witten curve. More precisely to have an ${\cal N}=1$
vacuum there must be some points on the quantum moduli space where
monopoles become massless, and at such points the corresponding
Seiberg-Witten curve becomes degenerate. The degeneration is such
that the Seiberg-Witten curve (\ref{G2q}) acquires two double
roots and two single roots. Having the locus of these
singularities one can read the quantum corrected moduli
parameters, $u$ and $v$ and thereby find the exact superpotential.
To have such a factorization we should impose the following
conditions
\be
{\partial {\cal P}_{\rm quan}(z,x)\over \partial z}|_{z_0,x_0}=0\;,\;\;\;\;\;
{\cal P}_{\rm quan}(z_0,x_0)=0\;,\;\;\;\;\;
{\partial {\cal P}_{\rm quan}(z,x)\over \partial z}|_{z_0,x_0}=0\;.
\ee
From the first condition one finds $z_0=\pm \Lambda^4/6$. And therefore
the other conditions read
\bea
&x_0^6-2ux_0^4+u^2x_0^2-v\mp 2
\Lambda^4(x_0^2-{1\over 3}u)=0\;,&
\cr &&\cr &
3x_0^4-4ux_0^2+u^2\mp 2
\Lambda^4=0\;.&
\label{Conds}
\eea
Now the task is to minimize the superpotential subject to the
above conditions. A standard procedure is to introduce Lagrange
multipliers
\bea
W&=&g_2u-g_6v+{4\over 27}g_6u^3+A\left(3x_0^4-4ux_0^2+u^2\mp 2
\Lambda^4\right)\cr &&\cr
&+&B\left(x_0^6-2ux_0^4+u^2x_0^2-v\mp 2
\Lambda^4(x_0^2-{1\over 3}u)\right)\;.
\eea
From the equation of motion for $x_0$ one finds $A=0$, while from the
equation of motion for $v$ one gets $B=-g_6$. Finally the equation of
motion for $u$ leads to the following condition
\be
3x_0^3-3ux_0^2+{4\over 3}u^2\mp\Lambda^4+{3g_2\over 2g_6}=0\;,
\ee
which together with (\ref{Conds}) can be used to find $u,v$ and $x_0$ as
the following
\be
u=3{ e}^2\mp{\Lambda^4\over 2{e}^2}\;,\;\;\;\;\;
v=\mp4e^2\Lambda^4+{\Lambda^8\over 3{ e}^2}\mp{\Lambda^{12}\over
54e^6}\;,\;\;\;\;\;x_0=3e^2\mp{\Lambda^4\over 6e^2}\;,
\label{yyy}
\ee
where $e=(-g_2/4g_6)^{1/4}$. Therefore the curve is factorized as follows
\be
x^6-2ux^4+u^2x^2-v\mp 2
\Lambda^4(x^2-{1\over 3}u)=(x^2-3{ e}^2\pm{\Lambda^4\over 6{ e}^2})^2\;
(x^2\pm{2\Lambda^4\over 3{ e}^2})\;.
\ee
Setting $\Lambda=0$ one recognizes this solution as the case where the
gauge group is classically broken into $SU(2)\times U(1)$.
Moreover the quantum superpotential can be obtained
from the expression of $u$ and $v$ which is given by
\bea
W&=&g_2u+g_6(-v+{4\over 27} u^3)\cr
&=&3g_2e^2+4g_6e^6\pm 2\sqrt{-g_2g_6}\Lambda^4\;,
\eea
in agreement with the field theory result \cite{Landsteiner:1996ut}.

To study the effective superpotential for the case where
classically the gauge group is not broken, we should look for
a factorization of the Seiberg-Witten curve in such a way that
in the $\Lambda\rightarrow 0$ the curve behaves as
${\cal P}\sim x^6$. To find such a solution we note that the most
general factorization would be as follows
\be
x^6-2ux^4+u^2x^2-v\mp 2
\Lambda^4(x^2-{1\over 3}u)=\left(x^2-x_{\mp}^2\right)^2
\left(x^2-y_{\mp}^2\right)\;,
\ee
which leads to the following conditions
\bea
&x_-^6-2ux_-^4+u^2x_-^2-v- 2
\Lambda^4(x_-^2-{1\over 3}u)=0\;,&
\cr &&\cr
&x_+^6-2ux_+^4+u^2x_+^2-v+ 2
\Lambda^4(x_+^2-{1\over 3}u)=0\;,&
\cr &&\cr
&3x_-^4-4ux_-^2+u^2- 2
\Lambda^4=0\;,\cr &&\cr
&3x_+^4-4ux_+^2+u^2+ 2
\Lambda^4=0\;,&
\label{ooo}
\eea
which can be solved for $x_{\pm},u$ and $v$. The result is
\be
u=3^{1/4}2\Lambda^2,\;\;\;\;\;v={4\over 3^{1/4}}\;\Lambda^6\;,\;\;\;\;\;
x_{\mp}^2={\sqrt{3}\mp 1\over 3^{1/4}}\Lambda^2\;.
\label{zzz}
\ee
The curve is then factorized as
\be
x^6-2ux^4+u^2x^2-v\mp 2
\Lambda^4(x^2-{1\over 3}u)=\left(x^2-{\sqrt{3}\mp 1\over 3^{1/4}}
\Lambda^2\right)^2 \left(x^2-{2(\sqrt{3}\pm 1)\over 3^{1/4}}
\Lambda^2\right)\;.
\ee
It is now obvious to see that this solution corresponds to the situation
where classically the gauge group remains unbroken.

It is also interesting to note that the whole Seiberg-Witten curve is also
factorized in this case as follows
\be
(3\pm2\sqrt{3})(z+{\Lambda^8\over 36 z})+x^4-{2(3\pm\sqrt{3})\over 3^{3/4}}
\Lambda^2 x^2\pm {3\pm2\sqrt{3}\over 3}\Lambda^4=0\;,
\ee
which should capture the information of the low energy theory.

Finally the effective superpotential reads
\bea
W&=&g_2u-g_6v+{4\over 27}g_6u^3\cr &&\cr
&=&4 \left({3^{1/4}\over 2}g_2\Lambda^2-{3^{3/4}\over 27}
g_6\Lambda^6\right)\;.
\label{ppp}
\eea

We note that in comparison with the case where the gauge group is
classically broken to $SU(2)\times U(1)$ the four conditions
(\ref{ooo}) must be satisfied simultaneously, which means that
in this case four monopoles become massless. To see this manifestly
we note that there are two inequivalent moduli spaces for $G_2$ \footnote{
Having two copies of the moduli space for $G_2$ has also been noticed
in \cite{Landsteiner:1996ut}.}. In fact the condition
${\partial {\cal P}_{\rm quan}\over \partial z}=0$ besides the solution
we have been considering so far, $z_0=\pm\Lambda^4/6$, has another
solution which is given by the following algebraic equation
\be
(z+{\Lambda^8\over 36 z})=x^4-{u\over 3}x^2\;,
\ee
which generates a polynomial of eighth order
\be
P_8=12x^8-12ux^6+4u^2x^4-3vx^2+\Lambda^8\;.
\ee
Therefore the conditions for having a degenerate curve are now given by
\be
P_8|_{x_0}=0\;,\;\;\;\;\;\;\;\;\;\;\;{\partial P_8\over \partial x}|_{x_0}=0\;.
\ee

One can now proceed to find the points on the moduli space where the curve
becomes degenerate. Doing so, we will get the same solution as before, namely
for the case where the gauge group is classically broken to $SU(2)\times
U(1)$ one finds
\be
u=3{ e}^2\mp{\Lambda^4\over 2{e}^2}\;,\;\;\;\;\;
v=\mp4e^2\Lambda^4+{\Lambda^8\over 3{ e}^2}\mp{\Lambda^{12}\over
54e^6}\;,
\ee
and the curve is factorized as
\be
12x^8-12ux^6+4u^2x^4-3vx^2+\Lambda^8=12\left(x^2\pm{\Lambda^4\over
6e^2}\right)^2
\left(x^4-3x^2e^2+3e^4\pm x^2{\Lambda^4\over e^2}\right)\;.
\ee

On the other hand for the situation where classically the gauge group
remains unbroken the solution for the parameters of the moduli space is
the same as before, {\it i.e.} $u=3^{1/4}2\Lambda^2,\;
v={4\over 3^{1/4}}\;\Lambda^6$, and the curve is factorized as
\be
12x^8-12ux^6+4u^2x^4-3vx^2+\Lambda^8=12
\left(x^2-{\sqrt{3}+1\over 2\times 3^{1/4}}\Lambda^2\right)^2
\left(x^2-{\sqrt{3}-1\over 2\times 3^{1/4}}\Lambda^2\right)^2\;.
\ee
As we see these two solutions give two different factorizations in this
branch. Indeed in the first case we get the points where two monopoles
become massless while in the second one we get the points where
four monoploes become massless.

One could also consider the points of the moduli space where some
mutually non-local monopoles become massless. These would lead to
a superconformal field theory \cite{{Argyres:1995jj},
{Argyres:1995xn}}. In the $G_2$ case from the first branch where
we get a polynomial of sixth order, there is only one way to get
such a point where the curve is factorized as
\be
x^6-2ux^4+u^2x^2-v\mp 2
\Lambda^4(x^2-{1\over 3}u)= (x^2-x_0^2)^3\;.
\ee
which means that one should impose the condition that the curve has
triplet roots. Doing so one gets the following solution for $u$ and $v$
\be
u=\sqrt{6}\;\Lambda^2\;,\;\;\;\;\;v={10\sqrt{6}\over 9}\;\Lambda^6\;,
\ee
and the curve is factorized only for plus sign as follows
\be
x^6-2ux^4+u^2x^2-v+ 2
\Lambda^4(x^2-{1\over 3}u)=
\left(x^2-{2\sqrt{6}\over 3}\;\Lambda^2\right)^3\;.
\ee

Doing same for the second branch one finds
\be
12x^8-12ux^6+4u^2x^4-3vx^2+\Lambda^8=12
\left(x^2-{\sqrt{6}\over 2}\Lambda^2\right)
\left(x^2-{\sqrt{6}\over 6}\Lambda^2\right)^3\;.
\ee
This case corresponds to the situation where the discriminant of the
quantum curve and its derivative with respect to $u$ and $v$ vanish
\cite{Landsteiner:1996ut}.

\section{Theory on $R^3\times S^1$}

Let us now consider the ${\cal N}=1$ $G_2$ supersymmetric gauge
theory on $R^3\times S^1$ space. This model can be thought of as
an ${\cal N}=2$ SYM theory on $R^3\times S^1$ deformed by a tree
level potential ${\rm Tr}W(\phi)$. To study the quantum theory one
may use the corresponding integrable model. Actually since the
moduli of the quantum curve parameterize the quantum moduli space
of the theory, one might suspect that ${\rm Tr}(M^k)$ is the
quantum corrected version of the classical gauge invariant
parameter ${\rm Tr}(\phi^k)$\footnote{Actually, the precise
statement is that the $z$-independent part of ${\rm Tr}(M^k)$ is
the quantum corrected version of ${\rm Tr}(\phi^k)$}. In other
words, in view of the proposal made in \cite{Boels:2003fh} the
effective superpotential should be ${\rm Tr}W(M)$. To be specific,
consider a deformation of the theory with the following tree level
superpotential
\be
W=g_2u+g_6(-v+{4\over 27}u^3)={1\over 4}g_2{\rm Tr}(\phi^2)-
{1\over 6}g_6{\rm Tr}(\phi^6)+{11\over 864}g_6{\rm Tr}(\phi^2)^3\;.
\ee
According to the proposal of \cite{Boels:2003fh} the effective
superpotential is given by
\bea
W&=&{1\over 4}g_2{\rm Tr}(M^2)-{1\over 6}g_6{\rm Tr}(M^6)
+{11\over 864}g_6{\rm Tr}(M^2)^3-
{5\over 2}g_6 {\rm Tr}(M^2)
(z+{y_0y_1^2y_2\over z})\cr &&\cr
&+&L\log\left({y_0y_1^2y_2\over
\Lambda^8/36}\right)\;.
\label{kkk}
\eea
Here we have also imposed the constraint on $y_i$ using a Lagrange
multiplier $L$.

\subsection{Case 1: Unbroken gauge symmetry}
To get a supersymmetric vacuum one needs to minimize this
superpotential with respect to $\phi_i$ and $y_i$. This can be
done using the equations of motion of the fields $\phi_i$ and
$y_i$. The corresponding equations are given in appendix A. One
can easily see that one solution to the equations is given by
\be
\phi_1=\phi_2=0,\;\;\;\;\;y_1={2\over 3}y,\;\;\;\;\;y_2={1\over 3}y,\;\;\;\;\;
y_0=y\;,
\ee
where $y=3^{1/4}\Lambda^2/2$. This solution corresponds to the situation
in which the gauge group is classically unbroken. Moreover the gauge invariant
parameters read $u=3^{1/4}2\Lambda^2$ and $v=4/3^{1/4}\Lambda^6$.
Plugging this solution into
the expression of the effective potential one finds
\be
W=4 \left({3^{1/4}\over 2}g_2\Lambda^2-{3^{3/4}\over 27}
g_6\Lambda^6\right)\;,
\ee
in agreement with (\ref{ppp}).

This result can be used to study the four dimensional theory on
$R^4$. For example let us use this result to integrate in the
glueball field $S$ for the theory on $R^4$. To do this we note
that the Lagrange multiplier can be interpreted as the glueball
field. To integrate it in one needs to minimize the effective
superpotential without using its equation of motion for L, and we
get (replacing $L$ by $S$)
\be
W=4g_2y-{32\over 27}g_6y^3+S\log\left({3\Lambda^8\over 16y^4}\right)\;.
\ee
The next step is to integrate out $y$
\be
{\partial W\over \partial y}=4g_2-{32\over 9}\;g_6y^2-4{S\over y}=0\;,
\ee
which is an equation that can be used to solve for $y$. In fact
the solution can be given in power series of $S$. Up to ${\cal
O}(S^8)$ one finds
\be
y={1\over g_2}\;S+{8\over 9}\;{g_6\over g_2^4}\; S^3+{64\over 27}\;
{g_6^2\over g_2^7}\; S^5
+{2048\over 243}\;{g_6^3\over g_2^{10}}\; S^7\;.
\ee
Plugging the above expression for $y$ into the effective superpotential
one gets
\be
W=-4S\left(\log\left({2S\over 3^{1/4}g_2\Lambda^2}\right)-1\right)
-{32\over 27}\;{g_6\over g_2^3}\;S^3-{128\over 81}\;{g_6^2\over
g_2^6}\;S^5 -{8192\over 2187}\;{g_6^3\over g_2^9}\;S^7\;.
\ee
up to order eight in the glueball field $S$.

\subsection{Case 2: $G_2$ broken to $SU(2)\times U(1)$}
The equations of motion coming from the potential (\ref{kkk}) also
have another solution. In fact it can also be seen that the
following ansatz solves the equations given in appendix A,
\be
\phi_1=\phi,\;\;\;\;\;\phi_2=2\phi,\;\;\;\;\;y_0=\pm{4\Lambda^4\over 9e^2},
\;\;\;\;\;y_1=-{3e^2\over 4},\;\;\;\;y_2=\pm{\Lambda^2\over 9 e^2},
\label{eq:solSU2U1}
\ee
where $e=(-g_2/4g_4)^{(1/4)}$ and $\phi$ is given by
\be
\phi^2=-{1\over 12}\left(e^2\pm{10\Lambda^4\over 27e^2}\right)\;.
\ee
Looking at the limit $\Lambda\rightarrow 0$, one can see that this
solution corresponds to the situation where the gauge group is
classically broken into $SU(2)\times U(1)$ (see (\ref{eq:clasSU2U1})).

The gauge invariant parameters are found to be
\be
u=3e^2\mp{\Lambda^4\over 2e^2},\;\;\;\;\;
v=\mp 4e^2\Lambda^4+{\Lambda^8\over 3e^2}\mp {\Lambda^{12}\over 54e^6}\;,
\ee
and therefore the quantum superpotential reads
\be
W=3g_2 e^2+4g_6e^6\pm 2\sqrt{-g_2g_6}\;\Lambda^4\;,
\ee
which is the same as the one we found in the field theory.

Since quantum mechanically the gauge symmetry is broken to $U(1)$,
we expect a free parameter in the solution. Note that the
situation differs from the $U(N)$ case, there we had a center of
mass $U(1)$ that did not manifest itself in the solution. Because
$G_2$ is a simple Lie group we do not have this center of mass
$U(1)$. Thus we would expect to see one free parameter in our
solution. Obviously the solution (\ref{eq:solSU2U1}) does not have
a free parameter, which means that the ansatz $\phi_2=2\phi_1 =
2\phi$ somehow fixes this parameter. So the solution we have found
is merely a special case in a one-parameter family of solutions.

In fact the situation is very similar to the ${\cal N}=1\; SU(3)$
case where the gauge group is classically broken into $SU(2)\times
U(1)$. Similarly to the $G_2$ model, in the IR limit the $SU(2)$
factor of remaining gauge group gets confined and we are left with
only $U(1)$. Therefore one would expect to get a one-parameter
family of solutions. To see this let us consider the ${\cal N
}=1\; SU(3)$ case on $R^3\times S^1$ in more detail (see also
\cite{Boels:2003fh}). Consider the tree level superpotential
\be
W={g_2\over 2}\;{\rm Tr}(\phi^2)+{g_3\over 3}\;{\rm Tr}(\phi^3)\;,
\ee
where $\phi$ is the adjoint scalar. The quantum superpotential is then given
by
\be
W={g_2\over 2}\;{\rm Tr}(M^2)+{g_3\over 3}\;{\rm Tr}(M^3)\;,
\ee
with
\be
M=\pmatrix{\phi_1& y_1& z\cr 1& \phi_2 & y_2\cr {y_0\over z}& 1& -\phi_1-\phi_2}
\ee
being the Lax matrix of the corresponding integrable model.

One can write down the equations for the extrema of $W$ and then solve it.
For the situation we are interested in, where the gauge group is
classically broken into $SU(2)\times U(1)$ one could start from an
ansatz in which $\phi_1=\phi,\;\phi_2=-2\phi$ and solve the equation.
Doing so one finds
\be
\phi={g_2\over g_3}\pm{\Lambda^2\over \phi_0},\;\;\;\;\;\;y_0=
{\Lambda^4\over \phi_0^2},\;\;\;\;\;\;y_1=\phi_0\Lambda,\;\;\;\;\;\;
y_2=\phi_0\Lambda,
\ee
where $\phi_0$ is a solution of the following equation
\be
g_3\phi_0^3\pm 3g_2\Lambda\phi_o+2g_3\Lambda^3=0\;.
\ee

On the other hand relaxing the condition $\phi_2=-2\phi_1$ one can
also find other solutions, namely
\be
\phi_1={g_2\over g_3}\pm{\Lambda^3\over y},\;\;\;\;
\phi_2={-2g_2\over g_3}\mp{\Lambda^3\over y}\mp {1\over \phi_0},\;\;\;\;
y_0={\Lambda^3\over y\phi_0},\;\;\;\;y_1=\phi_0\Lambda^3,\;\;\;\;
y_2=y,
\ee
where $\phi_0$ is a solution of the following equation
\be
g_3y^2\phi_0^2+(3g_2y\pm g_3\Lambda^3)\phi_0+g_3y=0\;,
\ee
which is a one-parameter solution, as expected. Therefore the solution with
$\phi_2=-2\phi_1$ is merely a special case in a one-parameter family of
solutions. Nevertheless we note, however, that both of these solutions
lead to the same superpotential which is
\be
W={g_2^3\over g_3^2}\pm 2g_3\Lambda^3\;.
\ee
Obviously, the free parameter corresponds to a flat direction, and
as far as the superpotential is concerned having one special
solution is enough to determine the value of the superpotential in
the supersymmetric vacua.

Nevertheless we can still see the existence of the free parameter
by considering flows in the integrable system whose Lax matrix we
are using in constructing the quantum superpotential. In general
there are two independent flows, generated by $M^2$ and $M^6$, or
equivalently by $u$ and $v$. The flows of the integrable system
act on the Lax matrix via commutators
\begin{equation}
    \frac{\partial}{\partial t_k} M = [ M, {\cal L} ( M^{k-1} ) ],
\end{equation}
but one can also consider the flows of the dynamical variables
$\xi \in \{\phi_i,y_j\}$ by calculating the Poisson brackets
\begin{equation}
    F_k(\xi) = \frac{\partial}{\partial t_k} \xi = \{ \xi, \mathrm{Tr} M^k \}.
\end{equation}
To calculate the Poisson brackets one has to identify the coordinates
that correspond to the conjugate momenta $\phi_i$, these are the $x$'s
appearing in
\begin{equation}
    y_i = \exp ( \alpha_i \cdot x),
\end{equation}
with the $\alpha_i$ the simple roots of $D^{(3)}_4$
\begin{equation}
    \alpha_0 = -(2 \alpha_1 + \alpha_2),\quad \alpha_1 =
(0,\sqrt{2}),\quad \alpha_2 = (\frac{1}{2}\sqrt{6},-\frac{3}{2}
\sqrt{2}). \label{eq:g2roots}
\end{equation}
The Poisson brackets then read
\begin{eqnarray}
    \{ \phi_i , y_j \} = (\alpha_j)_i y_j ,
\end{eqnarray}
where $(\alpha_j)_i$ is the $i$-th component of the $j$-th root in the
basis (\ref{eq:g2roots}). Using these brackets it is straightforward to
calculate the flows $F_k(\xi)$. We find that for all $\xi \in
\{\phi_i,y_j\}$ the two flows $F_2$ and $F_6$ are
related in the following way:
\begin{equation}
     F_6(\xi) = -\frac{9}{4} (3 \frac{m}{g} + 176 \Lambda^4 + 2112
\Lambda^8 \frac{g}{m}) F_2(\xi).
\end{equation}
So, indeed, there is exactly one independent flow and therefore precisely
one free parameter in our solution. This establishes the fact that the
symmetry is broken down to a single $U(1)$.

\section{Factorization}

\subsection{Deriving the resolvent for $G_2$}

In this section we present some preliminary results that are a first step
towards generalizing the algebraic geometric proof of the factorization
of Seiberg-Witten curves given in section 4 of \cite{Boels:2003at}. The
results will be applied to $G_2$.
A more detailed treatment will be given in a future publication.

It is a well-known fact that the Seiberg-Witten curve has an
underlying integrable system. The integrable system is
characterized by the existence of a complete set of action-angle
variables. In terms of these variables the evolution in the
phase-space of the classical mechanical system becomes quite
simple, half of the variables are conserved and the other half
(the angle variables) evolve with constant velocity. Further, to
the integrable system one can associate a Riemann surface, which
is equivalent to the Seiberg-Witten curve. The conserved
quantities then correspond to the moduli of this surface and the
angle variables are coordinates on the Jacobian of this Riemann
surface\footnote{In general, only a subset of the moduli
correspond to action variables, and the number of flows need
therefore not be equal to the dimension of the Jacobian.}. The
equations of motion of the integrable system correspond to linear
flows on the Jacobian.

The main idea of section 4 of \cite{Boels:2003at} is that the
superpotential is at an extremum if the velocities of the flows on the
Jacobian are zero. The velocities of the flows are expressed in terms of
the superpotential $W(x)$ and the one-forms $\omega_k$
(see \cite{kacmoerbeke}):
\begin{equation} \label{flowv}
    v_k(W) = \mathrm{res}_{x=\infty} \left ( W'(x) \omega_k \right).
\end{equation}
Let us also remind the reader that we can express the quantum
superpotential as a residue
\begin{displaymath}
    W_\mathrm{quantum} = \mathrm{res}_{x=\infty} \left ( W(x) R(x) \right),
\end{displaymath}
with $R(x)$ the gauge theory resolvent,
\be R(x) = {\rm Tr} \frac{1}{x-\Phi} . \ee

It turns out to be possible to express both the one-forms $\omega_k$ and
the resolvent in terms of a single function:
\begin{equation}
    \Omega(x,u_k) = (\log\det (x-M(z))) |_{z^0},
    \label{eq:defOmega}
\end{equation}
by which we mean the $z$-independent part of $\log\det (x-M(z))$.
Further, $M(z)$ is the Lax matrix (with spectral parameter $z$) of
the integrable system that underlies the Seiberg-Witten curve and
the $u_k$ are the moduli of this curve. As it stands, $\Omega$ is
not well defined, because we have to extract the $z$-independent
part of some complicated function with branch cuts. One way to
define $\Omega$ is as follows. Since the characteristic polynomial
$\det (x-M(z))$ is symmetric under the interchange of $z$ and
$1/z$, we can write
\begin{displaymath}
    \det(x-M(z)) = a_0^2 \prod_{t=1}^r (a_t-z)(a_t - 1/z)
\end{displaymath}
and this allows us to define
\begin{displaymath}
    \Omega \equiv 2 \sum_{t=0}^r \log a_t.
\end{displaymath}

Having defined $\Omega$, the resolvent is given by
\begin{equation}\label{eq:def_resolvent}
    R(x) = \partial_x \Omega(x, u_k)
\end{equation}
and the one-forms by
\begin{equation}\label{eq:def_oneforms}
    \omega_k = \frac{\partial \Omega}{\partial u_k}\mathrm{dx}
\end{equation}
Actually, the definition of $\Omega$ still suffers from minus sign
ambiguities, which will be fixed by demanding that the resolvent that
follows from this $\Omega$ has the expansion
\begin{equation}\label{eq:expresolv}
    R(x) = \sum_{i=1}^\infty \frac{\mathrm{tr}(M(z)^{i-1})|_{z^0}}{x^i}.
\end{equation}

In order to show that this proposal makes sense, we will calculate the
resolvent and one-forms for $U(N)$. From the curve for $U(N)$
\begin{displaymath}
    \det(x-M(z)) = P_N(x) + (-1)^N (z+\frac{1}{z})
\end{displaymath}
we easily derive that $a_0^2 = 1/a_1$ and
\begin{displaymath}
    a_1 = (P + \sqrt{P^2 - 4})/2.
\end{displaymath}
This yields for the function $\Omega$
\begin{displaymath}
    \Omega = \log((P + \sqrt{P^2 - 4})/2),
\end{displaymath}
from which we derive the usual resolvent
\begin{equation}
    R(x) = \partial_x \Omega = \frac{P'(x)}{\sqrt{P(x)^2 - 4}}
\end{equation}
and one-forms
\begin{equation}
\omega_k = \partial_{u_k} \Omega  \mathrm{dx}=
\frac{x^{N-k}}{\sqrt{P(x)^2 -4}}\mathrm{dx}
\end{equation}

In order to apply this procedure to $G_2$ we must first compute the roots
$a_t$ for the $G_2$ curve. The algebraic curve is given by
\begin{equation}
3 \left ( z - \frac{\Lambda^8}{36 z} \right ) ^2 - x^2 \left ( z +
\frac{\Lambda^8}{36 z} \right ) ( 6 x^2 -2 u) - x^2 P(x) =0
\end{equation}
written in terms of $y = z + \frac{\Lambda^8}{36 z}$ this reads
\begin{equation}
    3 y^2 - x^2 y ( 6 x^2 - 2 u) - x^2 P(x) - \frac{\Lambda^{8}}{3} = 0.
\end{equation}
This equation has two solutions:
\begin{equation}
    y_\pm = x^2 \left ( x^2 - \frac{u}{3} \right ) \pm \frac{1}{3}
\sqrt{x^4 (3 x^2 -u)^2 + 3 x^2 P(x) + \Lambda^{8}},
\end{equation}
yielding the four roots of the algebraic curve
\begin{equation}
    z_{+\pm} = \frac{1}{2} y_+ \pm \frac{1}{6} \sqrt{9 y^2_+ -\Lambda^8},
\qquad
    z_{-\pm} = \frac{1}{2} y_- \pm \frac{1}{6} \sqrt{9 y^2_- -\Lambda^8}.
\end{equation}
To write down $\Omega$ we have to make a choice for the roots. One should
pick one root from $\{ z_{++}, z_{+-}\}$ and one from
$\{z_{--},z_{-+}\}$, so there are four possible choices:
\begin{eqnarray}
\Omega &=& \eta (\log z_{++} + \epsilon \log z_{--}), \mbox{ with}
\epsilon^2 = \eta^2  = 1 \nonumber \\
    &=& \eta ( \log \left ( \frac{1}{2} y_+ + \frac{1}{6} \sqrt{9 y^2_+
- \Lambda^8}\right ) + \epsilon \log \left (\frac{1}{2} y_-
+\frac{1}{6} \sqrt{9 y^2_- - \Lambda^8} \right ) )
\end{eqnarray}
If we choose $\eta=1, \epsilon=-1$ the resolvent reads
\begin{equation}
    R(x) = \partial_x \Omega = \frac{ 3 \partial_x y_+ }{\sqrt{9 y_+^2
- \Lambda^8}} - \frac{3\partial_x y_-}{\sqrt{9 y_-^2 - \Lambda^8}} .
\end{equation}
The expansion of the resolvent around $x=\infty$ should have the form of
(\ref{eq:expresolv}). Indeed, when we do the expansion we get
\begin{eqnarray}
    R(x) &=& \frac{8}{x} + \frac{4u}{x^3} +  \frac{4 u^2}{x^5} +
 \frac{4 u^3 + 6 v}{x^7} +  \frac{4 u^4 + 16 uv + \frac{20
\Lambda^8}{3}}{x^9} \nonumber \\
     &+& \frac{4 u^5 + 30 u^2 v + 30 u \Lambda^8}{x^{11}} + \frac{4
u^6 + 48 u^3 v + 6 v^2 + \frac{250}{3}\Lambda^8}{x^{13}}
    + \mathcal{O}\left (\frac{1}{x^{15}}\right ).
\end{eqnarray}
One can check that the coefficients in this expansion correspond
to the traces of powers of the Lax matrix. Classically, one could
write for the resolvent :$P_6'(x)/P_6(x)$, this would generate the
correct expansion if one would set to zero the terms in the
expansion that explicitly depend on $\Lambda$. Apparently this
naive guess for the resolvent is correct up to order $1/x^7$.

This resolvent also hints at the existence of a hyper-elliptic
curve for $G_2$. This can be seen as follows. The resolvent can be
written in the form
\begin{equation}
    R(x) = \frac{r(x)}{x^2 \sqrt{P_6(x)^2 - 4 \Lambda^8 (x^2- u/3)^2}}
\end{equation}
with $r(x)$ some function without poles. Comparing this resolvent to that
of $U(N)$ leads us to suggest that
\begin{equation}
    y^2 = P_6(x)^2 - 4 \Lambda^8 (x^2- u/3)^2
\end{equation}
is in fact a hyper-elliptic curve for $G_2$. Indeed, in analyzing the
factorization of the $G_2$ curve, expressions like $P_6(x) \pm 2
\Lambda^4 (x^2-u/3)$ pop up everywhere.

Notice that the resolvent of the gauge theory contains arbitrarily
high powers of the adjoint scalar field. The precise definition of
such operators in the quantum theory depends on a choice of UV
completion of the theory. The integrable system prefers one
particular UV completion, which is the one where we define
\be \mathrm{tr}(\Phi^{i})\equiv
 \mathrm{tr}(M(z)^{i})|_{z^0} .
\ee
In the case of $U(N)$, this was also the UV completion preferred
by string theory. We see that the integrable system provides a
natural UV completion for the exceptional gauge groups as well. It
would be interesting to explore other UV completions, e.g. those
obtained by taking the Lax matrix in another representation, but
we leave that for future work.

We now want to use the flow equations (\ref{flowv}) to determine
the minima of the superpotential. One therefore has to calculate
the one-forms $\partial_{u,v} \Omega$
\begin{eqnarray}
\omega_u = \partial_u \Omega = \partial_u a(x) R_1(x) + \partial_ub(x)
R_2 (x)\\
\omega_v = \partial_v \Omega = \partial_v a(x) R_1(x) + \partial_v b(x) R_2 (x).
\end{eqnarray}
The conditions that the flows on the Jacobian vanish ($v_l=0$) then imply:
\begin{eqnarray}\label{eq:jabflow}
x^2 R_2(x) W'(x) = r_v(x) + \sum_{l=1}\frac{c_l}{x^{2l+1}} \nonumber \\
\left ( -\frac{x^2}{3} R_1(x) + 4 x^4 (2u - 3 x^2) R_2(x) \right )
W'(x) = r_u(x) + \sum_{l=1}\frac{d_l}{x^{2l+1}}.
\end{eqnarray}
The flow equations for the $U(N)$ case allowed us to derive the
factorization of the gauge theory and Matrix model curve (see
\cite{Boels:2003at} section 4), in a similar spirit equations
($\ref{eq:jabflow}$) should somehow define the analogue of the
Matrix model curve for $G_2$. Unfortunately, we have not yet
succeeded in writing (\ref{eq:jabflow}) in a more manageable form,
and it is therefore harder to draw general conclusions from these
equations. In the next section we will use an alternative method
to work out the factorization of the Matrix model curve for a
superpotential with terms up to order six.

In the remaining part of this section we will show that
(\ref{eq:jabflow}) is indeed equivalent to minimizing the superpotential.
For definiteness we will choose the superpotential to be $W'(x) = g_2 x +
g_6 x^5$. The values of $u$ and $v$ in the minimum
determine the Seiberg-Witten curve completely and therefore also the
factorization properties of this curve. To study the factorization
properties it is useful to consider the conditions for the curve to
develop a double zero:
\begin{eqnarray}
    P_6(x_0) &=& \pm 2 \Lambda^4 (x_0^2 - u/3) \label{eq:factcond} \\
    P_6'(x_0) &=& \pm 4 \Lambda^4 x_0,
\end{eqnarray}
these equations can be used to solve for $u$ and $v$ in terms of $x_0$.
So there is only one free parameter, not two. Therefore the two equations
(\ref{eq:jabflow}) are replaced by the single equation
\begin{equation}
    \label{eq:eomx0}
    \mathrm{res}_{x=\infty} \left ( \partial_{x_0}\Omega W'(x) \right )= 0.
\end{equation}
This equation can be used to solve for $x_0$, which will allow us to
express $u$ and $v$ in terms of the coupling constants and the energy
scale $\Lambda$. One can then substitute $u$ and $v$ into equation
(\ref{eq:factcond}) and study its factorization p
roperties. For the superpotential $W'(x) = g_2 x + g_6 x^5$ we find three
classes of solutions (note that we consider single trace operators here,
so these results should not be compared with the results from the
previous sections)
\begin{enumerate}
 \item $x_0 = 0 \Rightarrow P_6(x) - 2 \Lambda^4 (x^2 - u/3) = x^4 ( x^2 -
2 \sqrt{2} \Lambda^2)$
 \item $x_0 = \eta \left (\frac{8}{3}\right )^{1/4},\quad
 \eta^4=-1\Rightarrow P_6(x) - 2 \Lambda^4 (x^2 - u/3) = (x^2 \pm 2
\mathrm{i}   \sqrt{\frac{2}{3}}\Lambda^2 )^3$ \\\\This solution is similar
to the superconformal solution.
     \item $x_0 = \epsilon \left ( \Lambda^4 - 6 e - \frac{5}{33}
 \sqrt{\frac{9}{e^2}- 66 \frac{\Lambda^4}{e}+ 22 \Lambda^8 } \right
 )^{1/4},\quad \epsilon^4=1, \quad e = \frac{g_6}{g_2}$ \\\\$\Rightarrow
P_6(x) - 2 \Lambda^4 (x^2 - u/3) = (x^2 - \alpha)(x^2
 - \beta)^2$ \\\\ Here $\alpha$ and $\beta$ are some (messy) expressions
in $e$ and $\Lambda$.
\end{enumerate}
 In order to check the claim that equation (\ref{eq:eomx0}) is equivalent
to minimizing the superpotential, if suffices to minimize
\begin{equation}
    W = \frac{g_2}{2} u + \frac{g_6}{6} v + A(P_6(x_0)  \pm 2
\Lambda^4 (x_0^2 - u/3)) + B (P_6'(x_0)  \pm 4 \Lambda^4 x_0)
\end{equation}
 with respect to $u, v, A, B$ and $x_0$. The calculations are pretty
straightforward and we find complete agreement, suggesting that
$\Omega$ indeed generates the one-forms as described.

\subsection{Extremization problem: Proof of $B_{l-1}^2F_{12-2l}={W^\prime}(x)^2+f_8(x)$}
In order to understand the curve factorization better we
apply the same analysis as in \cite{Cachazo:2001jy} to the $G_2$ case. We
consider
a single trace superpotential and for the matrix $\Phi$ we consider three
independent fields $\phi_1,\phi_2,\phi_3$ as the non zero diagonal
components. We know that these three fields are not in fact independent and classically
 \be
 \phi_3=\phi_2-\phi_1\;.
 \ee
 At the quantum level the following constraint holds
 \be
 u_4=({u_2\over 2})^2\;.
 \ee
 We impose this constraint by a Lagrange multiplier, $C$.
 The effective superpotential will read
 \bea
  W_{eff}&=&\sum_{r=1}^{3}g_{2r}u_{2r}+
[L_i(P_6(x)-2\epsilon_i\Lambda^4(x^2-{u_2\over6}))|_{x=p_i}+
Q_i({\partial\over \partial x}P_6
  -4\epsilon_i\Lambda^4x)|_{x=p_i}]
  \cr&+&C(u_4-({u_2\over2})^2)\;,
  \label{pot}
  \eea
where $l$  is the number of double zeroes, $\epsilon$ is a second
root of unity and the $p_i$ are the points where the factorization
occurs. In lines parallel to \cite{Cachazo:2001jy} one can see
that $Q_i=0$ and 
\be P_6=\langle
det(xI-\Phi)\rangle=\sum_{j=0}^{\infty}x^{6-j}s_l|_+\;, 
\ee 
where ``+'' means the polynomial part of the series. Using the relation,
${\partial s_j\over \partial u_k}=-s_{j-k}$ and upon variation of
(\ref{pot}) with respect to all $u_r$ one finds 
\bea
&g_2&=\sum_{i=1}^{l}L_i[\sum_{j=0}^{6}p_i^{6-j}
s_{j-2}-\epsilon_i{\Lambda^4\over 3}]+C{u_2\over2}
\cr&g_4&=\sum_{i=1}^{l}\sum_{j=0}^{6}L_ip_i^{6-j}s_{j-4}-C
\cr&g_6&=\sum_{i=1}^{l}L_i 
\eea 
Multiplying (\ref{pot}) by
$x^{2r-1}$ and summing over $r$ and imposing the $L_i$ constraints
one will find 
\bea W^\prime(x)&=&\sum_{r=1}^{3}g_{2r}x^{2r-1}
\cr&=&\sum_{r=-\infty}^{3}\sum_{i=1}^{l}
\sum_{j=0}^{6}x^{2r-1}p_i^{6-j}s_{j-2r}L_i
-\sum_{i=1}^{l}2L_i\epsilon_i\Lambda^4x^{-1}
(p_i^2-{u_2\over6})\cr&+&C({u_2\over2}x-x^3)
-L{\Lambda^4\over3}x+{\cal O}(x^{-2})
\cr&=&\sum_{i=1}^{l}{P_6(x;<u>)\over {x-p_i}}L_i
-\sum_{i=1}^{l}2L_i\epsilon_i\Lambda^4x^{-1}
(p_i^2-{u_2\over6})+C({u_2\over2}x-x^3)
\cr&-&L{\Lambda^4\over3}x+{\cal O}(x^{-2})\;, 
\eea 
where $L\equiv\sum_{i=1}^{l}L_i\epsilon_i$. Defining $B_{l-1}(x)$ by 
\be
\sum_{i=1}^{l}{L_i\over{x-p_i}}={B_{l-1}(x)\over H_l(x)} 
\ee 
one has 
\bea 
W^\prime(x)&+&\sum_{i=1}^{l}2L_i\epsilon_i\Lambda^4x^{-1}
(p_i^2-{u_2\over6})-C({u_2\over2}x-x^3)
+L{\Lambda^4\over3}x\cr&=&B_{l-1}(x)\sqrt{F_{12-2l}(x)
+{4\Lambda^8(x^2-{u_2\over6})^2\over H_l(x)^2}}+{\cal
O}(x^{-2})\;. \label{res} 
\eea Squaring (\ref{res}) we find
\be \label{finres}
B_{l-1}^2F_{12-2l}={W^\prime}(x)^2+2g_6Cx^8+{\cal O}(x^6)\;, 
\ee
which is the desired result. This suggests that the right hand
side is somehow related to the matrix model curve and that
therefore the appropriate matrix model curve may well be
hyperelliptic, just as we found hints that the gauge theory curve
may also be represented in hyperelliptic form, as we discuss in
the next section.

\section{Hyperelliptic curve for $G_2$}

Our considerations in section 5 about the resolvent for $G_2$
suggest that the exact result for ${\cal N}=2$ $G_2$ SYM theory
can be obtained from a hyperelliptic curve given by 
\be
y^2=(x^6-2ux^4+u^2x^2-v)^2-4\Lambda^8(x^2-{u\over 3})^2\;. 
\label{hc}
\ee

Actually having a hyperelliptic curve for $G_2$ was first
suggested in \cite{Danielsson:1995zi} though the proposed curve
leads to incorrect singularities of the moduli space
\cite{Landsteiner:1996ut}. Therefore it was believed that the
correct curve for $G_2$ which comes from the integrable model need
not be hyperelliptic. The corresponding hyperelliptic curve is
\cite{Danielsson:1995zi} 
\be
y^2=(x^6-2ux^4+u^2x^2-v)^2-4\Lambda^8x^4\;. 
\label{wrong} \ee 
As we
see the hyperelliptic curve (\ref{hc}) suggested by the resolvent
of $G_2$ is a simple modification of this curve, though this
simple modification leads to completely different physics.

To study the singularity structure of the hyperelliptic curve let
us consider the case where the gauge group is classically broken
to $SU(2)\times U(1)$. From the field theory considerations we
know that the quantum corrections to the gauge invariant variables
are given by (\ref{yyy}). Upon eliminating $e$ one finds 
\be
\Delta_{\mp}^{\rm field\;th.}=\pm 12u^3v\mp 81v^2-108vu\Lambda^4
+16\Lambda^4u^4\pm 12u^2\Lambda^8+96\Lambda^{12}=0\;, 
\label{dyon}
\ee 
as the condition for a vacuum with a massless dyon. Note that
$\Delta_{+}^{\rm field\;th.}$ and $\Delta_{-}^{\rm field\;th.}$
intersect transversally  in four points (see equation (\ref{zzz}))
\be 
(u,v)=(e^{{in\pi\over 2}}3^{1/4}2\Lambda^2,-e^{{3in\pi\over
2}}{4\over 3^{1/4}} \Lambda^6),\;\;\;\;\;\;\;{\rm
for}\;\;n=0,1,2,3\;, 
\ee 
which is equal to number of the
supersymmetric ground states of ${\cal N}=1\;G_2$ SYM theory.

Let us now consider the hyperelliptic curve (\ref{wrong}). The
discriminant of the curve is given by 
\be
\Delta^{h}=v\Delta^h_+\Delta^h_- 
\ee 
with 
\be
\Delta^h_{\pm}=27v^2-4vu^3\mp72uv\Lambda^4\pm8u^4\Lambda^4+
32u^2\Lambda^8\pm32\Lambda^{12}\;. 
\ee 
It can be shown
\cite{Landsteiner:1996ut} that this leads to an incorrect number
of ${\cal N}=1$ vacua  and moreover the overall factor of $v$ in
the discriminant gives a monodromy which is not present in the
Weyl group of $G_2$. Therefore one might conclude that the
hyperelliptic curve (\ref{wrong}) is not a proper curve describing
${\cal N}=2\;G_2$ SYM theory.

On the other hand the discriminant of the hyperelliptic curve
(\ref{hc}) is given by 
\be
\Delta=(4u^2\Lambda^8-9v^2)(-4u^3+27v)\Delta^{\rm field\; th.}_+
\Delta^{\rm field\; th.}_-\;. 
\ee 
In comparison with the field
theory result the discriminant has two extra overall factors. The
first one was already present in the $G_2$ curve coming from the
integrable system (\ref{G2q}) which is believed to be an
accidental singularity \cite{Landsteiner:1996ut}. The second one,
$-4u^3+27v$, is present in our hyperelliptic curve. Nevertheless
since $\Delta^{\rm field\; th.}_{\pm}$ in the discriminant of the
curve (\ref{hc}) coincide precisely with the gauge theory condition
for having massless dyons, (\ref{dyon}), one might suspect that this
singularity is accidental too. Moreover, by construction, 
the curve also gives the correct factorization. Therefore
one could believe  that the curve (\ref{hc}) is a proper curve
describing ${\cal N}=2\; G_2$ SYM theory. Definitely this issue
deserves to be studied more carefully.

\section{Conclusion}
We have studied ${\cal N}=2\; G_2$ SYM theory deformed by a tree
level superpotential on $R^3\times S^1$ using the corresponding
integrable model of the theory which is the periodic Toda lattice
based on dual Affine $G_2$. For the cases where the gauge group is
classically broken and where it remains unbroken we have obtained
the vacuum structure and the exact superpotential of the theory
both by conventional field theory methods (Seiberg-Witten curve
factorization) and by integrable model techniques (using the Lax
operator) and we have shown complete agreement between the two
approaches in each case.

We have also put forward a general recipe for deriving the
resolvent and the one-forms from the Seiberg-Witten curve. This
method was applied to the Seiberg-Witten curve for $G_2$. Using
the one-forms to calculate the flows on the Jacobian we reproduced
the conditions for an extremum of the superpotential, as the flow
equations should. Also, the proposed resolvent appears to be
correct, since it reproduces the correct expansion around
$x=\infty$. The resolvent obtained from the Lax matrix provides a
natural UV completion of the theory, and we also found a set of
algebraic equations that somehow encode the appropriate notion of
a matrix model curve for the gauge group $G_2$. Clearly, more work
is needed to determine the precise structure of this generalized
matrix model curve.

The extremization problem has also been considered with a proof
allowing us to state the problem in purely algebraic terms.
Contradicting earlier beliefs we have presented evidences and
arguments supporting the existence of a hyperelliptic curve for
$G_2$. This last suggestion deserves further study which we postpone
to future work.\\

{\bf Acknowledgments: } JdB and JW thank the Stichting FOM for
support. JW thanks Rutger Boels and Robert Duivenvoorden for
useful discussions. JdB would also like to thank the Aspen Center
for Physics where this work was completed.

\newpage
\appendix
\section{Equations of motion}

Here we present the equations of motion for the potential given in
(\ref{kkk}). For $\phi_1$ we get

\bea
0&=&3g_2(2\phi_1-\phi_2)+2 g_6 \bigg{(}-27 y_2 y_1\phi_2+9y_1y_0\phi_2
-12y_2y_0\phi_2+{4\over 3}y_0^2\phi_2-12y_0\phi_2^3\cr &&\cr
&+&9y_0\phi_1^2\phi_2
-81\phi_2^3\phi_1^2+54\phi_2^2\phi_1^3
+{1\over 3}y_0^2\phi_1+
6y_0\phi_1\phi_2^2+27y_2^2\phi_1+54y_1\phi_2^2\phi_1\cr &&\cr
&-&81y_2\phi_1^2\phi_2+54y_2\phi_1\phi_2^2-6y_2y_0\phi_1
-27y_1\phi_2^3+27\phi_2^4\phi_1\bigg{)}\;.
\eea
For $\phi_2$ one has
\bea
0&=&3g_2(2\phi_2-\phi_1)+2g_6\bigg{(}-{8\over 3}y_0^2\phi_2
+27y_1^2\phi_2-27y_2\phi_1y_1+9y_0\phi_1y_1
-12y_2y_0\phi_1\cr &&\cr
&+&{4\over 3}y_0^2\phi_1-36y_0\phi_1\phi_2^2
+3y_0\phi_1^3-81\phi_2^2\phi_1^3+27\phi_2\phi_1^4
+6y_0\phi_1^2\phi_2+24y_0\phi_2^3+54y_1\phi_2\phi_1^2\cr &&\cr
&-&27y_2\phi_1^3+54y_2\phi_1^2\phi_2+24y_2y_0\phi_2
-81y_1\phi_2^2\phi_1-12y_1y_0\phi_2+54\phi_2^3\phi_1^2\bigg{)}\;.
\eea
For $y_0$ one finds
\bea
{L\over y_0}&=&g_2+g_6\bigg{(}-{16\over 3}y_0\phi_2^2+{4\over 9}y_0^2
+{8\over 3}y_1y_0+18y_1\phi_1\phi_2-24\phi_1\phi_2y_2
-12y_1y_2\cr &&\cr
&+&{16\over 3}\phi_1y_0\phi_2-24\phi_2^3\phi_1
+6\phi_1^3\phi_2+{2\over 3}y_0\phi_1^2+6\phi_2^2\phi_1^2
+12\phi_2^4+12y_2^2-6\phi_1^2y_2\cr &&\cr
&+&24\phi_2^2y_2
-12y_1\phi_2^2-{16\over 3}y_2y_0\bigg{)}\;.
\eea
For $y_1$ one gets
\bea
{2L\over y_1}&=&3g_2+g_6\bigg{(}54y_1\phi_2^2+{4\over 3}y_0^2
-54\phi_1\phi_2y_2+18\phi_1y_0\phi_2-12y_2y_0
+54\phi_2^2\phi_1^2\cr &&\cr
&-&54\phi_2^3\phi_1-12y_0\phi_2^2\bigg{)}\;.
\eea
For $y_2$ one gets
\bea
{L\over y_2}&=&3g_2+g_6\bigg{(}-54y_1\phi_1\phi_2-24\phi_1y_0\phi_2
-12y_1y_0+54\phi_1^2y_2+24y_2y_0-54\phi_1^3\phi_2\cr &&\cr
&+&54\phi_2^2\phi_1^2-6y_0\phi_1^2+24y_0\phi_2^2-{8\over 3}y_0^2\bigg{)}\;,
\eea
and finally for $L$ we get $y_0y_1^2y_2=\Lambda^8/36$.


\begin{thebibliography}{99}

\bibitem{Intriligator:1995au}
K.~A.~Intriligator and N.~Seiberg,
``Lectures on supersymmetric gauge theories and electric-magnetic
duality,''
Nucl.\ Phys.\ Proc.\ Suppl.\  {\bf 45BC}, 1 (1996)
[arXiv:hep-th/9509066].

\bibitem{Dijkgraaf:2002dh}
R.~Dijkgraaf and C.~Vafa, ``A perturbative window into
non-perturbative physics,'' arXiv:hep-th/0208048.

\bibitem{Bershadsky:1993cx}
M.~Bershadsky, S.~Cecotti, H.~Ooguri and C.~Vafa,
``Kodaira-Spencer theory of gravity and exact results for quantum
string amplitudes,''
Commun.\ Math.\ Phys.\  {\bf 165}, 311 (1994)
[arXiv:hep-th/9309140].

\bibitem{Cachazo:2001jy}
F.~Cachazo, K.~A.~Intriligator and C.~Vafa,
``A large N duality via a geometric transition,''
Nucl.\ Phys.\ B {\bf 603}, 3 (2001)
[arXiv:hep-th/0103067].

\bibitem{Cachazo:2002pr}
F.~Cachazo and C.~Vafa,
``N = 1 and N = 2 geometry from fluxes,''
arXiv:hep-th/0206017.


\bibitem{Dijkgraaf:2002fc}
R.~Dijkgraaf and C.~Vafa,
``Matrix models, topological strings, and supersymmetric gauge theories,''
Nucl.\ Phys.\ B {\bf 644}, 3 (2002)
[arXiv:hep-th/0206255].

\bibitem{Dijkgraaf:2002vw}
R.~Dijkgraaf and C.~Vafa,
``On geometry and matrix models,''
Nucl.\ Phys.\ B {\bf 644}, 21 (2002)
[arXiv:hep-th/0207106].

\bibitem{Dijkgraaf:2002xd}
R.~Dijkgraaf, M.~T.~Grisaru, C.~S.~Lam, C.~Vafa and D.~Zanon,
``Perturbative computation of glueball superpotentials,''
arXiv:hep-th/0211017.

\bibitem{Cachazo:2002ry}
F.~Cachazo, M.~R.~Douglas, N.~Seiberg and E.~Witten,
``Chiral rings and anomalies in supersymmetric gauge theory,''
JHEP {\bf 0212}, 071 (2002)
[arXiv:hep-th/0211170].

\bibitem{Seiberg:1996nz}
N.~Seiberg and E.~Witten,
``Gauge dynamics and compactification to three dimensions,''
arXiv:hep-th/9607163.

\bibitem{Katz:1996th}
S.~Katz and C.~Vafa,
``Geometric engineering of N = 1 quantum field theories,''
Nucl.\ Phys.\ B {\bf 497}, 196 (1997)
[arXiv:hep-th/9611090].

\bibitem{Gomez:1997uh}
C.~Gomez and R.~Hernandez,
``M and F-theory instantons, ${\cal N}=1$ supersymmetry and fractional
topological charge,''
Int.\ J.\ Mod.\ Phys.\ A {\bf 12}, 5141 (1997)
[arXiv:hep-th/9701150].

\bibitem{Lee:1997vp}
K.~M.~Lee and P.~Yi,
``Monopoles and instantons on partially compactified D-branes,''
Phys.\ Rev.\ D {\bf 56}, 3711 (1997)
[arXiv:hep-th/9702107].

\bibitem{Aharony:1997bx}
O.~Aharony, A.~Hanany, K.~A.~Intriligator, N.~Seiberg and M.~J.~Strassler,
``Aspects of N = 2 supersymmetric gauge theories in three dimensions,''
Nucl.\ Phys.\ B {\bf 499}, 67 (1997)
[arXiv:hep-th/9703110].




\bibitem{Boels:2003fh}
R.~Boels, J.~de Boer, R.~Duivenvoorden and J.~Wijnhout,
``Nonperturbative superpotentials and compactification to
three dimensions,''
arXiv:hep-th/0304061.

\bibitem{Dorey:1999sj}
N.~Dorey,
``An elliptic superpotential for softly broken ${\cal N}=4$
supersymmetric  Yang-Mills theory,''
JHEP {\bf 9907}, 021 (1999)
[arXiv:hep-th/9906011].

N.~Dorey, T.~J.~Hollowood and S.~Prem Kumar,
``An exact elliptic superpotential for ${\cal N}=1^*$
deformations of finite  ${\cal N}=2$ gauge theories,''
Nucl.\ Phys.\ B {\bf 624}, 95 (2002)
[arXiv:hep-th/0108221].

\bibitem{Boels:2003at}
R.~Boels, J.~de Boer, R.~Duivenvoorden and J.~Wijnhout,
``Factorization of Seiberg-Witten curves and compactification to three
dimensions,''
arXiv:hep-th/0305189.

\bibitem{Hollowood:2003ds}
T.~J.~Hollowood,
``Critical points of glueball superpotentials and equilibria of
integrable systems,''
arXiv:hep-th/0305023.


\bibitem{Alishahiha:2003pu}
M.~Alishahiha and A.~E.~Mosaffa,
``On effective superpotentials and compactification to three dimensions,''
JHEP {\bf 0305}, 064 (2003)
[arXiv:hep-th/0304247].


\bibitem{Gorsky:1995zq}
A.~Gorsky, I.~Krichever, A.~Marshakov, A.~Mironov and A.~Morozov,
``Integrability and Seiberg-Witten exact solution,''
Phys.\ Lett.\ B {\bf 355}, 466 (1995)
[arXiv:hep-th/9505035].

\bibitem{Martinec:1995by}
E.~J.~Martinec and N.~P.~Warner,
``Integrable systems and supersymmetric gauge theory,''
Nucl.\ Phys.\ B {\bf 459}, 97 (1996)
[arXiv:hep-th/9509161].

\bibitem{Nakatsu:1995bz}
T.~Nakatsu and K.~Takasaki,
``Whitham-Toda hierarchy and ${\cal N}=2$ supersymmetric Yang-Mills theory,''
Mod.\ Phys.\ Lett.\ A {\bf 11}, 157 (1996)
[arXiv:hep-th/9509162].


\bibitem{Donagi:1995cf}
R.~Donagi and E.~Witten,
``Supersymmetric Yang-Mills Theory And Integrable Systems,''
Nucl.\ Phys.\ B {\bf 460}, 299 (1996)
[arXiv:hep-th/9510101].

\bibitem{Eguchi:1995jh}
T.~Eguchi and S.~K.~Yang,
``Prepotentials of $N=2$ Supersymmetric Gauge Theories and Soliton Equations,''
Mod.\ Phys.\ Lett.\ A {\bf 11}, 131 (1996)
[arXiv:hep-th/9510183].

\bibitem{Itoyama:1995nv}
H.~Itoyama and A.~Morozov,
``Integrability and Seiberg-Witten Theory: Curves and Periods,''
Nucl.\ Phys.\ B {\bf 477}, 855 (1996)
[arXiv:hep-th/9511126].

\bibitem{D'Hoker:1999ft}
E.~D'Hoker and D.~H.~Phong,
``Lectures on supersymmetric Yang-Mills theory and integrable systems,''
arXiv:hep-th/9912271.

\bibitem{Mironov:2000se}
A.~Mironov,
``Seiberg-Witten theories, integrable models and perturbative
prepotentials,''
arXiv:hep-th/0010078.


\bibitem{Itoyama:2002rk}
H.~Itoyama and A.~Morozov,
``Calculating gluino condensate prepotential,''
Prog.\ Theor.\ Phys.\  {\bf 109}, 433 (2003)
[arXiv:hep-th/0212032].

H.~Itoyama and A.~Morozov,
``Gluino-condensate (CIV-DV) prepotential from its Whitham-time
derivatives,''
arXiv:hep-th/0301136.



\bibitem{Aganagic:2003xq}
M.~Aganagic, K.~Intriligator, C.~Vafa and N.~P.~Warner,
``The glueball superpotential,''
arXiv:hep-th/0304271.


\bibitem{Landsteiner:1996ut}
K.~Landsteiner, J.~M.~Pierre and S.~B.~Giddings,
``On the moduli space of ${\cal N}=2$ supersymmetric
$G_2$ gauge theory,''
Phys.\ Rev.\ D {\bf 55}, 2367 (1997)
[arXiv:hep-th/9609059].


\bibitem{Pesando:1995bq}
I.~Pesando,
``Exact results for the supersymmetric $G_2$ gauge theories,''
Mod.\ Phys.\ Lett.\ A {\bf 10}, 1871 (1995)
[arXiv:hep-th/9506139].

S.~B.~Giddings and J.~M.~Pierre,
``Some exact results in supersymmetric theories based on exceptional groups,''
Phys.\ Rev.\ D {\bf 52}, 6065 (1995)
[arXiv:hep-th/9506196].

P.~Pouliot,
``Chiral duals of nonchiral SUSY gauge theories,''
Phys.\ Lett.\ B {\bf 359}, 108 (1995)
[arXiv:hep-th/9507018].


\bibitem{Brandhuber:2003va}
A.~Brandhuber, H.~Ita, H.~Nieder, Y.~Oz and C.~Romelsberger,
``Chiral rings, superpotentials and the vacuum structure of
${\cal N}=1$  supersymmetric gauge theories,''
arXiv:hep-th/0303001.




\bibitem{vilenkin}
N.~Ja.~Vilenkin and A.~U.~Klimyk, ``Representation of Lie Groups and
Special Functions,'' Volume 3,
Kluwer (1992)

\bibitem{Argyres:1995jj}
P.~C.~Argyres and M.~R.~Douglas,
``New phenomena in SU(3) supersymmetric gauge theory,''
Nucl.\ Phys.\ B {\bf 448}, 93 (1995)
[arXiv:hep-th/9505062].

\bibitem{Argyres:1995xn}
P.~C.~Argyres, M.~Ronen Plesser, N.~Seiberg and E.~Witten,
``New N=2 Superconformal Field Theories in Four Dimensions,''
Nucl.\ Phys.\ B {\bf 461}, 71 (1996)
[arXiv:hep-th/9511154].

T.~Eguchi, K.~Hori, K.~Ito and S.~K.~Yang,
``Study of $N=2$ Superconformal Field Theories in $4$ Dimensions,''
Nucl.\ Phys.\ B {\bf 471}, 430 (1996)
[arXiv:hep-th/9603002].

\bibitem{kacmoerbeke}
P.~van Moerbeke and D.~Mumford,
``The spectrum of difference operators and algebraic curves'',
Acta Math. 143, 1-2, 93--154 (1979),


\bibitem{Danielsson:1995zi}
U.~H.~Danielsson and B.~Sundborg,
``Exceptional Equivalences in ${\cal N}=2$ Supersymmetric Yang-Mills Theory,''
Phys.\ Lett.\ B {\bf 370}, 83 (1996)
[arXiv:hep-th/9511180].

M.~Alishahiha, F.~Ardalan and F.~Mansouri,
``The Moduli Space of the Supersymmetric $G_{2}$ Yang-Mills Theory,''
Phys.\ Lett.\ B {\bf 381}, 446 (1996)
[arXiv:hep-th/9512005].

M.~R.~Abolhasani, M.~Alishahiha and A.~M.~Ghezelbash,
``The moduli space and monodromies of the ${\cal N}=2$ supersymmetric
Yang-Mills  theory with any Lie gauge groups,''
Nucl.\ Phys.\ B {\bf 480}, 279 (1996)
[arXiv:hep-th/9606043].

%



\end{thebibliography}
\end{document}